\documentstyle[11pt,paspconf,epsf]{article}

\begin{document}

\title{Radio, Infrared and X-ray observations of GRS 1915+105}
\author{R. P. Fender}
\affil{Astronomy Centre, University of Sussex, Brighton BN1 9QH U.K.}
\author{G. G. Pooley}
\affil{MRAO, Cavendish Laboratory, Cambridge CB3 0HE, U.K.}
\author{C. R. Robinson, B. A. Harmon, S. N. Zhang}
\affil{MSFC, Huntsville, AL 35812, U.S.A.}
\author{C. Canosa}
\affil{Astronomy Centre, University of Sussex, Brighton BN1 9QH U.K.}
\begin{abstract}

We present multiwavelength observations of the superluminal jet 
source GRS 1915+105 in 1996 April--May, over which period a variety of
phenomena, including radio QPO, strong infrared emission lines and rapid
X-ray flickering and outbursts were observed.

\end{abstract}

\keywords{GRS 1915+105,radio,infrared,X-ray}

\section{Introduction}

GRS 1915+105 is an energetic X-ray transient with
associated relativistic jets (e.g. Mirabel \& Rodriguez 1994).
The source undergoes recurrent outbursts
with correlated radio -- X-ray
behaviour (Foster et al 1996). There is no optical counterpart
but spectroscopy of a variable infrared counterpart has revealed
HI \& HeI emission lines during a period of outburst (Castro-Tirada
et al 1996). 

GRS 1915+105 is now being monitored in the radio, soft \& hard X-ray
regimes by the Ryle Telescope (RT), XTE/ASM and GRO/BATSE
respectively. We combine these data sets for the period 1996 April--May,
during which a deep infrared spectrum of the source was also obtained. 

\section{Results}

Fig 1 presents the RT, ASM and BATSE monitoring of
GRS 1915+105 over the period 1996 April--May, indicating the date
on which we obtained our infrared spectrum and when radio QPO were
observed. Lack of space precludes a discussion; we only summarise
the behaviour of the source in each energy regime.

\begin{itemize}
\item{{\bf Radio :}
GRS 1915+105 remained below detection levels ($\sim 0.5$ mJy) with
the RT until May 23 when it underwent a rapid flare
event. Between May 23--27 the source exhibited radio QPO with periods
in range 20 -- 100 min.}
\item{{\bf Infrared :}
the UKIRT IR spectrum obtained revealed (at least) strong 
HeI 2.06 $\mu$m and HI 2.16 $\mu$m emission.}
\item{{\bf Soft X-ray :}
GRS 1915+105 had been in a gradual decline until $\sim$ May 20,
when it began brightening.
During the period of the radio QPO many large
amplitude, rapid brightness variations were observed (see
also Greiner this proceedings).}
\item{{\bf Hard X-ray :}
the hard X-ray brightness of GRS 1915+105 varied over the entire
1996 April--May period, but shows activity around the period of the
radio flare, with a steady increase in flux up to $\sim 27$ May,
when the flux again declined.}
\end{itemize}

\begin{figure}
\plotone{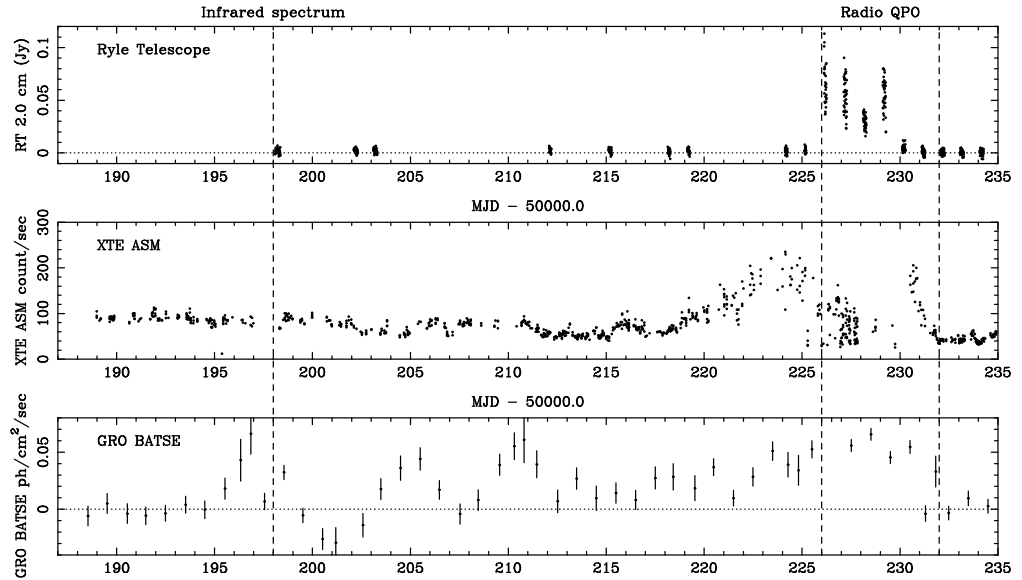}
\caption{Radio, XTE (ASM) and GRO (BATSE) monitoring of GRS 1915+105
over the period April--May 1996, indicating when our IR spectrum
was obtained and when radio QPO were observed.}
\end{figure}

\acknowledgments
Thanks to ASM RXTE team for quick-look results.

\end{document}